\documentclass[showpacs,preprintnumbers,amsmath,amssymb,prd]{revtex4}

\usepackage{graphicx}
\usepackage{dcolumn}
\usepackage{bm}

\newcommand{\mpc}{\, {\rm Mpc}}

\newcommand{\hmpc}{\, h^{-1} \mpc}
\newcommand{\ihmpc}{\, h\, {\rm Mpc}^{-1}}

\newcommand{\lyaf}{Ly$\alpha$ forest}
\newcommand{\bF}{\bar{F}}

\newcommand{\lr}{\lambda_{{\rm rest}}}
\newcommand{\lrmin}{\lambda_{{\rm rest, min}}}
\newcommand{\lrmax}{\lambda_{{\rm rest, max}}}
\newcommand{\lmin}{\lambda_{{\rm min}}}
\newcommand{\lmax}{\lambda_{{\rm max}}}

\newcommand{\vtheta}{{\mathbf \theta}}
\newcommand{\vk}{{\mathbf k}}
\newcommand{\vx}{{\mathbf x}}
\newcommand{\vdelta}{{\mathbf \delta}}
\newcommand{\vp}{{\mathbf p}}

\newcommand{\fid}{{\rm fid}}

\begin{document}

\title{ Dark energy and curvature from a future baryonic acoustic oscillation 
survey using the \lyaf\ }

\author{Patrick McDonald}
\email{pmcdonal@cita.utoronto.ca}
\affiliation{Canadian Institute for Theoretical Astrophysics, University of
Toronto, Toronto, ON M5S 3H8, Canada}

\author{Daniel Eisenstein}
\affiliation{Steward Observatory, University of Arizona, 933 N. Cherry Ave., 
Tucson, AZ 85121}

\date{\today}

\begin{abstract}

We explore the requirements for a Lyman-$\alpha$ forest survey designed to 
measure the angular diameter distance and Hubble parameter at $2\lesssim z
\lesssim 4$ using the standard ruler provided by baryonic acoustic oscillations
(BAO).  The goal would be to obtain a high enough density of sources to probe 
the three-dimensional density field on the scale of the BAO feature.  A 
percent-level measurement in this redshift range can almost double the Dark 
Energy Task Force Figure of Merit, relative to the case with only a similar 
precision measurement at $z\sim 1$, if the Universe is not assumed to be flat.
This improvement is greater than the one obtained by doubling the size of the 
$z\sim 1$ survey, with Planck and a weak SDSS-like $z=0.3$ BAO measurement 
assumed in each case.  Galaxy BAO surveys at $z\sim 1$ may be able to make an 
effective \lyaf\ measurement simultaneously at minimal added cost, because the 
required number density of quasars is relatively small.  We discuss the 
constraining power as a function of area, magnitude limit (density of quasars),
resolution, and signal-to-noise of the spectra.  For example, a survey covering
2000 sq. deg. and achieving $S/N=1.8$ per \AA\ at $g=23$ ($\sim 40$ quasars per
sq. deg.) with an $R\gtrsim 250$ spectrograph is sufficient to measure both the 
radial and transverse oscillation scales to 1.4\% from the \lyaf\ (or better, 
if fainter magnitudes and possibly Lyman-break galaxies can be used).  At fixed
integration time and in the sky-noise-dominated limit, a wider, noisier survey 
is generally more efficient; the only fundamental upper limit on noise being 
the need to identify a quasar and find a redshift.  Because the \lyaf\ is much 
closer to linear and generally better understood than galaxies, systematic 
errors are even less likely to be a problem.  

\end{abstract}


\maketitle

\section{Introduction}

The existence of new physics that causes the Universe to accelerate at late 
times is well established, both by observations of Type Ia supernovas 
\citep{2003ApJ...598..102K,2004ApJ...607..665R,2006A&A...447...31A} 
and by combinations of other 
observables \citep{2006astro.ph..4335S}.  The focus now is on probing the 
properties of the acceleration, most commonly thought to be caused by dark 
energy, a substance we know almost nothing about except that it must have
negative pressure.  We often parameterize dark energy by the equation of state,
$w=p/\rho$, where $p$ is the pressure and $\rho$ the density
($w=-1$ for a cosmological constant).  In general, $w$ can be time-dependent
\citep{1988PhRvD..37.3406R}. 

Acoustic oscillations before recombination
lead to a feature in the matter correlation 
function at the scale of the sound horizon at decoupling
\citep{1970ApJ...162..815P,1970Ap&SS...7....3S, 1978SvA....22..523D}.
CMB observations pin this scale at $143\pm 4$ Mpc 
\citep{2003ApJS..148..233P}.  
Recently, \cite{2005ApJ...633..560E} detected the expected enhancement 
(at $>3\sigma$) in 
the correlation of the Sloan Digital Sky Survey (SDSS) Luminous Red Galaxies 
(LRGs).  In coming years, precision measurements of the BAO feature will 
be used as a standard ruler to probe the equation of state of dark energy 
\citep{
1998ApJ...504L..57E, 1998ApJ...496..605E,
2001ApJ...557L...7C,
2003astro.ph..1623E,2003ApJ...594..665B,2003PhRvD..68h3504L,2003ApJ...598..720S,
2004ApJ...615..573M,
2005ApJ...631....1G,2005astro.ph..7457G,2005MNRAS.357..429A,2005MNRAS.363.1329B,
2006MNRAS.365..255B,2006MNRAS.366..884D}. 

We define the usual cosmological parameters in addition to $w$:
The scale factor is $a=1/(1+z)$, with Hubble parameter $H(z)=\dot{a}/a$ with 
$H(0)=H_0=h~100~{\rm km~s^{-1} Mpc^{-1}}$.  The baryon density is 
$\omega_b=\Omega_b h^2$ where $\Omega_b=\rho_b/\rho_c$ is the fraction of the
critical density, $\rho_c=3 H_0^2/8 \pi G$, in baryons.  The matter density is 
$\omega_m=\Omega_m h^2$, where $\Omega_m$ is the fraction of the
critical density in baryons plus cold dark matter.  The fraction of the 
critical density in dark energy is $\Omega_w$.  The curvature is
parameterized by $\Omega_k=1-\Omega_m-\Omega_w$.
The primordial power spectrum is parameterized by its amplitude $A$ and 
power law index, $n_s$ (we will not use tensors or running of the spectral 
index in this paper).  The Thompson
scattering optical depth to the CMB surface of last scattering is $\tau$.
We will sometimes use $\theta_s$
at the CMB last scattering redshift, defined below in Eq.~(\ref{thetas}), 
as a free parameter in place of $h$. 
We measure $\theta_s$ in degrees so a typical value is 0.6.

In general, the quantities measured most directly in a BAO study are
\begin{equation}
v_s(a)=H(a)~s~a
\end{equation}
in the radial direction and
\begin{equation} 
\theta_s(a)=\frac{s~a}{D_A(a)}
\label{thetas}
\end{equation}
in the transverse direction, where $s$ is the comoving sound horizon at 
decoupling
\citep{2005ASPC..339..215H}, and $D_A(a)$ is the angular diameter 
distance.  
For flat models, $D_A(a)/a=r(a)$ and $\theta_s(a)=s/r(a)$ with
\begin{equation}
r(a)=\int_a^1 
\frac{c}{a^{\prime 2}H(a^\prime)}da^\prime~.
\end{equation}
For curved models,
\begin{equation}
(1+z)~D_A(a)=R~\sinh\left[\frac{r(a)}{R}\right]
\end{equation}
with 
\begin{equation}
R=\frac{c}{H_0 \sqrt{\Omega_k}}~.
\end{equation}
Finally, 
\begin{equation}
s=\int_0^{a_r} \frac{c_s(a)}{a^2 H(a)} da
\end{equation}
where $a_r\simeq 1/1090$ is the recombination redshift (\citep{2005ASPC..339..215H}
gives a useful fitting formula for $a_r$'s relatively small parameter 
dependence), and $c_s(a)=c/\sqrt{3\left(1+3 \rho_b/ 4 \rho_\gamma\right)}$, 
where $\rho_\gamma$ is the photon density. 

In this paper, we discuss the possibility of measuring the BAO feature in a
\lyaf\ survey (see also \cite{2003dmci.confE..18W}). 
The acoustic scale $143~{\rm Mpc}$ corresponds to $\sim 1.3$ degrees on 
the sky at $z\sim 2.8$, our typical mean redshift.
On this large scale, the clustering of the
absorption can be assumed to follow a linear bias model in a way 
analogous to galaxies.  This has been studied in simulations by 
\cite{2003ApJ...585...34M}.  
The primary difference from galaxy surveys is in how the density field is 
sampled on relatively small scales.
With galaxies, we observe a collection of discrete points, which we believe are
a Poisson sampling of the underlying density field.  For the \lyaf, we observe 
a collection of skewers through the density field, obtaining a detailed picture
along the skewers, but no information in between.  The reader who is
uncomfortable with this should consider that they are probably perfectly 
comfortable constructing a density field from the ultimately zero dimensional
galaxy sampling -- in the forest we not only have a fully one-dimensional 
field, we even
measure a continuous value rather than the discrete presence or absence
of a galaxy.  For galaxies there is intrinsic noise due to the Poisson 
sampling, while for the \lyaf\ there is intrinsic aliasing-like noise due to 
the discrete transverse sampling.  As we will show, if the sampling of the 
\lyaf\ was only zero dimensional, this noise would take approximately the same 
white (uncorrelated) form as the galaxy noise.  Of course, the \lyaf\ also has 
the advantage of being sensitive to the near-mean-density intergalactic
medium (IGM), rather than the highly non-linear positions of galaxies, so we 
can more 
confidently back up our assumptions with near-first-principles calculations.
This is not to say that the \lyaf\ is decisively superior to galaxies as a 
high-z BAO probe.  Ultimately, both should work, and the resource requirements
to conduct an equivalent survey are the most important factor in deciding which
is better.  

Computing the 
expected errors obtainable from the \lyaf\ as a function of survey parameters
is the main aim of this paper.  In \S \ref{secsetup} we describe our basic 
method for estimating obtainable errors on $\theta_s$ and $v_s$.  
Then in \S \ref{secresults} we give the results.  
Finally, in \S \ref{seccosmology}, we discuss the usefulness for constraining
dark energy and curvature of a measurement of $\theta_s(z\sim 3)$ and
$v_s(z\sim 3)$. 

\section{ Setting up the calculation \label{secsetup}}

In this section we discuss our method for estimating the 
constraining power of future surveys.  First, 
in \S \ref{secdataconfig}, we define the parameters 
that describe the survey configuration.  Then, in \S
\ref{secfishmat}, we explain our Fisher matrix 
calculations.  Finally, in \S \ref{sectheorycor} we explain our theoretical 
model for the \lyaf\ power spectrum.

\subsection{Assumed data set \label{secdataconfig}}

We assume a square survey with area $A$ and comoving number
density of quasars $n_q$.
We usually ignore evolution across the redshift extent
of the survey, e.g., $n_q$ is the density at the central redshift.
One would of course need to consider evolution in an  analysis 
of real data, but, as we will show,
the resulting error bars will
not be increased relative to a Fisher matrix estimate that ignores
evolution.  $n_q$ will be a function of the central redshift and  magnitude 
limit of the survey, based on the luminosity function, $\frac{dn_q}{dm}(z)$,
taken from \cite{2006astro.ph..2569J}.
Beyond setting $n_q$, the luminosity function is relevant for setting the 
distribution of noise levels in spectra.
We assume that the survey probes the forest in the observed wavelength range 
$\lmin < \lambda < \lmax$,
which we translate to $z_{\rm min}$ and $z_{\rm max}$
using $\lambda = \lambda_\alpha (1+z)$ (with $\lambda_\alpha=1216$\AA).  
The central redshift
is just $(z_{\rm min}+z_{\rm max})/2$.
We assume the usable rest wavelength range in a single spectrum is
$\lrmin < \lr < \lrmax$, generally using $\lrmin=1041$\AA\ 
and $\lrmax=1185$\AA\ (this is very conservative - one could probably 
extend the baryonic oscillation analysis into the 
Lyman-$\beta$-influenced region without much difficulty).  
We will occasionally call the length
in the IGM probed by a single line of sight $L_q$ ($\sim 330 \hmpc$ for these 
limits).
The spectrograph is assumed to have rms 
resolution $\sigma_R$ (${\rm FWHM} = 2.355~\sigma_R = \lambda/R$).  
Spectral pixels are assumed to have full width $l_p=\sigma_R$.
We assume rms noise per pixel $\sigma_N$, 
in units of the mean transmitted flux level in the forest.

\subsection{ Fisher matrix calculation \label{secfishmat}}

We estimate the errors on parameters $p_i$, obtainable using a future 
survey, to be $(F^{-1})_{i i}$, where $F_{i j}$ is the Fisher matrix
\citep{1997ApJ...480...22T}:
\begin{equation}
F_{i j} = \left<\frac{\partial^2 \mathcal{L}}
{\partial p_i \partial p_j}\right>~,
\end{equation}
with 
\begin{equation}
2 \mathcal{L} = \ln \det \mathbf{C} + \mathbf{\delta}^T \mathbf{C}^{-1}
\mathbf{\delta}+ {\rm constant}
\end{equation}
for a real mean zero Gaussian vector $\vdelta$ of data points with covariance 
matrix 
$C_{ij}(\vp) = \left<\delta_i \delta_j\right>$.
From \cite{1997ApJ...480...22T} (and references therein), we have
\begin{equation}
F_{i j} = \frac{1}{2} {\rm Tr}\left(\mathbf{C}^{-1}
\frac{\partial\mathbf{C}}{\partial p_i} \mathbf{C}^{-1}
\frac{\partial\mathbf{C}}{\partial p_j}\right)~.
\label{eqFishmat}
\end{equation}
We could always take $\delta_i$ to be individual pixels in spectra and evaluate
Eq.~(\ref{eqFishmat}) by brute force; however, 
this calculation would quickly become difficult.  For example, our fiducial
survey would have about 40 million pixels.  While we could probably  
reduce this number enough to produce a reasonably invertible covariance matrix 
by scaling up from a much smaller survey and pushing the pixel size and angular
sampling to just barely resolve the BAO feature, it is better to look for a 
more efficient method.

Methods for applying the Fisher matrix formalism to a BAO survey using galaxies 
are well-developed \citep{2003ApJ...598..720S}.  Counts of 
galaxies (say, in cells smaller than the scale of interest) are the density
field of interest, but the calculations are done in Fourier space.  
Assuming an approximately uniform selection function,
the covariance matrix of Fourier modes is approximately diagonal if the 
wavenumber
$k_j$ is discretized in bands of width $2 \pi/L_j$, where $L_j$ is the survey   
width in direction $j$.
The covariance matrix of the real and imaginary parts of Fourier modes is 
simply 
$C_{ii}(\vp)=\frac{1}{2}\left[P_g(\vk_i,\vp)+P_N\right]$, 
where $P_g$ is the redshift-space galaxy power spectrum and 
$P_N=1/\bar{n}_g$ is 
the Poisson noise power for mean galaxy density $\bar{n}_g$.
Simple formulas for the Fisher matrix as an integral over $k$ can be derived.  

The \lyaf\ is slightly more complicated.  
Here, the quasar lines of sight going through some volume
are approximately random.  Our density estimate for the volume is not 
determined by the number of lines of sight probing it, but rather by some 
average of the absorption in those lines of sight.  
Working toward a covariance matrix of Fourier modes, 
we will take $\delta_i$ in the Fisher matrix to be the Fourier 
modes of the weighted density field
\begin{equation}
\delta(\vx) = \frac{w(\vx)}{\bar{w}}[\delta_F(\vx)+\delta_N(\vx)]~.
\end{equation}
where $\delta_F(\vx)=\delta_F(z,\vtheta)=F(z,\vtheta)/\bF-1$ where 
$F=\exp(-\tau)$ is the transmitted flux fraction in the forest, 
$\delta_N(\vx)$ is the spectral noise, $w(\vx)$ is the weight at $\vx$, and 
$\bar{w}$ is the mean weight.
We will not specify the weights at this point, but generally they should depend
on pixel noise variance. 
The relevant quantities are all defined over fully 
three-dimensional space -- if some location is not probed by a quasar line of
sight, the weight is simply zero.  
The correlation of Fourier modes of 
$\delta(\vx)$ is 
\begin{equation}
\left<\delta_\vk\delta_{\vk'}\right>=
\int d\vx~d\vx'\exp(i \vk\cdot\vx+i \vk'\cdot\vx')~
\left[\xi_F(\vx-\vx')+\xi_N(\vx,\vx')\right]
[1+\delta_w(\vx)][1+\delta_w(\vx')]~,
\end{equation}
defining $\delta_w(\vx)=w(\vx)/\bar{w}-1$, and 
$\xi_X\left(\vx,\vx'\right)=
\left<\delta_X(\vx)\delta_X(\vx')\right>$.  

We do not know the weights in advance, because they depend on 
the random quasar locations and luminosities.   
We can, however, evaluate the average of the weight term, which should give us
a good estimate of the constraining power of the survey (for a typical survey
of $\sim 10000$ quasars, a quantity measured from a single realization should 
not deviate much from the average quantity).  
Our pixel weights will be approximately uncorrelated in the transverse 
directions, and perfectly correlated in the radial direction (for 
simplicity we are assuming pixels in the same spectrum have the same noise). 
For a continuous white noise field, $\xi(\vx)=P~\delta^D(\vx)$, 
where $P$ is the constant power spectrum of the field.  
This gives 
\begin{equation}
\left<\delta_w(\vx)\delta_w(\vx')\right>=P_w^{\rm 2D} \delta^D(\vx_\perp) 
\end{equation}
where $P_w^{\rm 2D}$ is the power spectrum of $\delta_w$ in the two angular 
directions.
This means
\begin{equation}
\left<\delta_\vk\delta_{\vk'}\right>\simeq
(2\pi)^3 \delta^D(\vk+\vk') \left[P_F(\vk)+
P_F^{\rm 1D}(k_\parallel) P_w^{\rm 2D}+P^{\rm eff}_N\right]~,
\label{eqnewpower}
\end{equation}
where $P_F^{\rm 1D}(k_\parallel)$ is the usual 1D flux power spectrum 
measured along single lines of sight, and $P^{\rm eff}_N$ is the weighted 
noise power, which does not separate like the other term because the weights
depend on the noise amplitude.
Note that modes with different $\vk$ are still approximately uncorrelated.

We now need to evaluate the $P_w^{\rm 2D}$ term. 
The low-k power 
spectrum for a pixelated white noise field is $P=\sigma^2 v$, where $\sigma^2$
is the pixel variance and $v$ is the pixel volume (in a general sense, i.e., 
here the volume is actually an area because we are working in two dimensions).  
As a computational device, let us imagine pixelating the transverse directions
into small cells of angular width $l$ (we will eventually take 
$l\rightarrow 0$).  Then, $P_w^{\rm 2D}=\sigma_w^2 l^2$, with 
$\sigma^2_w=\left<w^2\right>/\bar{w}^2-1$.
The noise in a 
spectrum is a function of the apparent magnitude of the quasar, so the mean 
weight is 
\begin{equation}
\bar{w}=
\int_{-\infty}^{m_{\rm max}}dm \frac{dn_q}{dm}~L_q~l^2~ w(m)\equiv
l^2 L_q I_1~. 
\end{equation}
Note that $L_q l^2 dn_q/dm$ is the probability that a given small volume of 
the IGM 
will be probed by a line of sight to a quasar of apparent magnitude $m$.
Similarly,
\begin{equation}
\left<w^2\right>=
\int_{-\infty}^{m_{\rm max}} dm\frac{dn_q}{dm}~L_q~l^2~ w^2(m)\equiv
l^2 L_q I_2~. 
\end{equation}
Therefore, in the limit $l\rightarrow 0$, 
\begin{equation}
P^{\rm 2D}_w=\frac{I_2}{I_1^2 L_q}
\end{equation}
A similar calculation leads to 
\begin{equation}
P^{\rm eff}_N=\frac{I_3 l_p}{I_1^2 L_q}
\end{equation}
where
\begin{equation}
I_3\equiv \int_{-\infty}^{m_{\rm max}} dm\frac{dn_q}{dm}~\sigma_N^2(m) w^2(m)~.
\end{equation}
Note that $I_1 L_q/l_p$ is something like an effective 3D pixel density
(recall that $l_p$ is the pixel width), while 
$I_3/I_1$ is the effective noise variance in these pixels.

We take the weights to have the simple Feldman, Kaiser, \& Peacock (hereafter
FKP) inverse variance form \cite{1994ApJ...426...23F},
\begin{equation}
w(m)= \frac{P_S/P_N(m)}
{1+P_S/P_N(m)}~, 
\end{equation}
where $P_S$ is the typical signal power and $P_N(\vx)=
P_N[m(\vx)]$ is the noise power
level associated with the noise level at $\vx$ (if there is no quasar probing
$\vx$, this level is infinite so the weight is zero).  The numerator is chosen
to make $w(\vx)=1$ in the low noise limit.  
There is a subtlety at this point in 
that $P_N(m)$ is not unambiguously defined.  If quasars were evenly distributed
in angle, and all had the same magnitude, then clearly 
$P_N=\sigma_N^2 l_p/n_q L_q$.  The only reasonable possibility in the 
realistic case seems to be to take 
$P_N(m)=\sigma^2_N(m) l_p/ I_1 L_q=\sigma^2_N(m)/n_p^{\rm eff}$, where 
$n_p^{\rm eff}$ is the effective pixel density.  Note the similarity to galaxy
shot noise: if the overall mean density of galaxies is $\bar{\bar{n}}_g$,
and the mean at point $\vx$ is $\bar{n}_g(\vx)$, then the
Poisson-noise variance at point $\vx$ is 
$\sigma_P^2=\bar{\bar{n}}_g/\bar{n}_g(\vx)$ for 
cells with 
volume $1/\bar{\bar{n}}_g$, and the noise factor in the usual FKP weights is 
$\sigma^2_P(\vx)/\bar{\bar{n}}_g$  
(recall that our $m$ dependence is equivalent to $\vx$ dependence).
Since $n_p^{\rm eff}$ and the weights are mutually dependent, we determine
them using a few iterations.  After some experimentation showing that the 
results are not very sensitive to the choice of $P_S$, we take the 
shortcut of using one constant for $P_S$: the total flux power 
in the central model at $k=0.07 \ihmpc$, $\mu=k_\parallel/k=0.5$,
including the aliasing term (which is formally signal, 
although in practice it acts as noise).    

There are a couple of loose ends in this derivation.
First, we are going to assume that, having computed the power spectrum, i.e., 
the covariance matrix of Fourier
modes, averaged over realizations of the weights, we can simply proceed with 
the calculation of the constraining power of
a survey as if this were the exact power spectrum; however, the averages of 
the covariance matrix of power measurements, or of the Fisher matrix, are not 
necessarily the same as what one computes from the averaged power spectrum.
Ideally we would only average at the last possible point in the computation
(remember, the Fisher matrix is already an average of the derivatives of the 
likelihood function over all possible future data sets).  Unfortunately, 
without the assumption that the Fourier modes are uncorrelated, we are stuck
with the problem of inverting a very large covariance matrix, and now we would
need to do it many times to average over realizations.  
To investigate this problem, we computed the 
average covariance matrix of the power spectrum measurements and found that the
new terms are suppressed by a factor of the number of quasars
in the survey (i.e., the ratio of the area of the survey to the area per 
line-of-sight) relative to the standard 
terms, as long as the second term in Eq.~(\ref{eqnewpower}) is not much
larger than the first.  This will be the case any time the BAO measurement is
possible, so it seems we are safe to ignore this issue.  
Second, with no Nyquist frequency for a randomly 
sampled survey, it is not clear at what k we should stop counting modes.  This 
is worrisome because we will find that our constraining power will continue to 
increase somewhat 
beyond the transverse Nyquist frequency associated with the effective mean 
density of lines of sight.  This means that we are using more Fourier modes
than we have data points, and the assumption that they are uncorrelated 
probably has to break down somehow, e.g., due to an accumulation of small 
correlations.  To be conservative, we only use modes with
$k$ less than the Nyquist frequency associated with the effective mean density.
We also eliminate modes with $k>0.5\ihmpc$ (relevant only in the radial 
direction).  To allow for continuum fitting, we drop all the modes having the 
first $2 N_{\rm q, los}$ discrete values of $k_\parallel$, counting both 
positive and negative $k_\parallel$, where 
$N_{\rm q, los}$ is the number of spectra it takes to cover the redshift range
of our survey (the first 5 values of $k_\parallel$ for our standard survey). 
These last two cuts make no noticeable difference in the results.

We note for the future that one would probably perform the data analysis using 
a more sophisticated maximum likelihood method \citep{1997ApJ...480...22T}, but
this FKP-like method should be good enough for the Fisher matrix error 
estimate.  

\subsection{Three-dimensional \lyaf\ power spectrum \label{sectheorycor}}

Missing from the discussion so far has been any mention of how we
predict the \lyaf\ power spectrum and its parameter dependence that is needed
for the Fisher matrix calculation.  Fortunately, this is basically a solved 
problem:
We use the fitting formula for $P_F(\vk,\vp)$ from 
\cite{2003ApJ...585...34M}.  
We start with a standard $\Lambda$CDM transfer function from
CMBfast \citep{1996ApJ...469..437S} with 
$\sigma_8=0.897$, $\Omega_m=0.281$, $\Omega_b=0.0462$, $h=0.710$, 
and $n=0.980$ \citep{2005PhRvD..71j3515S}.  The amplitude and slope of the 
primordial 
power spectrum (i.e., $\sigma_8$ and $n$) are free parameters to be
marginalized over, while we leave the other parameters fixed because, from the
point of view of the \lyaf, they are either well constrained ($\Omega_b$) or 
degenerate with the slope and amplitude.    
Additionally, we marginalize over four parameters of the \lyaf\ model:  
the mean transmitted flux level, $\bF$, the temperature at the 
mean density, $T_{1.4}$, the slope of the power law temperature-density 
relation,
$\gamma-1$, and the large scale anisotropy parameter $\beta$
\citep{2003ApJ...585...34M}.  These 
parameters set the large scale bias of the \lyaf\ power spectrum (the bias
is effectively free because it is hyper-sensitive to $\bF$).  $\beta$ is 
given by \cite{2003ApJ...585...34M} as a function of the other parameters,
but we choose to marginalize over it to be sure we are not relying on this 
prediction.
Finally, the parameters of interest are the radial and angular scale factors,
$v_s(a)$ and $\theta_s(a)$,
which we use to convert the linear power spectrum from comoving Mpc/h units
to the observed velocity (redshift) and angular coordinates. 
There is little chance that our power spectrum predictions can be significantly
wrong, because they agree with high precision measurements of the 1D power in 
single quasar spectra 
\citep{2000ApJ...543....1M,2002ApJ...581...20C,2004MNRAS.347..355K,
2006ApJS..163...80M,2005ApJ...635..761M}.  
Note in particular that the aliasing-like noise
term in Eq.~(\ref{eqnewpower}), which turns out to be critical to our 
calculation, is precisely the well-measured 1D power. 

\section{Results \label{secresults}}

There are several interacting degrees of freedom for a survey so to give us a
concrete start we 
imagine the \lyaf\ survey piggy-backed onto something like the proposed WFMOS 
survey of low 
redshift galaxies \citep{2005astro.ph..7457G}.  This proposal is to observe 
two million galaxies in the redshift range $0.5<z<1.3$ in a 2000 sq. deg. area 
using an 8m telescope with an R=2000 spectrograph with exposures of 30 minutes.
We take the number density of quasars as a function of luminosity from 
\cite{2006astro.ph..2569J}, finding that 
g magnitude limits (21, 22, 23, 24, 25) correspond to (8, 20, 41, 77, 136) 
quasars per sq. deg. within the relevant redshift range.
We assume the noise is sky-dominated and for our baseline WFMOS-like survey 
assume S/N=(11, 4.5, 1.8, 0.7, 0.3) per 1\AA\ for 
g=(21, 22, 23, 24, 25). 
There are of course a substantial number of quasars brighter than any given
magnitude limit.  We track the S/N over the source counts,
rather than assuming that all quasars are at the limiting magnitude.
At $g\gtrsim 23$, the number density of Lyman-break galaxies (LBGs) starts to 
become high enough to provide significant extra sources.  
In cases where we include these, we take the LBG luminosity function from 
\cite{2000ApJ...544..218A}, assuming
g=R+0.7, finding (0.3, 116, 2325) LBGs per sq. deg. for $g<(23, 24, 25)$
($m_*^R=24.54$ for these galaxies). 

\begin{figure}
\resizebox{\textwidth}{!}{\includegraphics{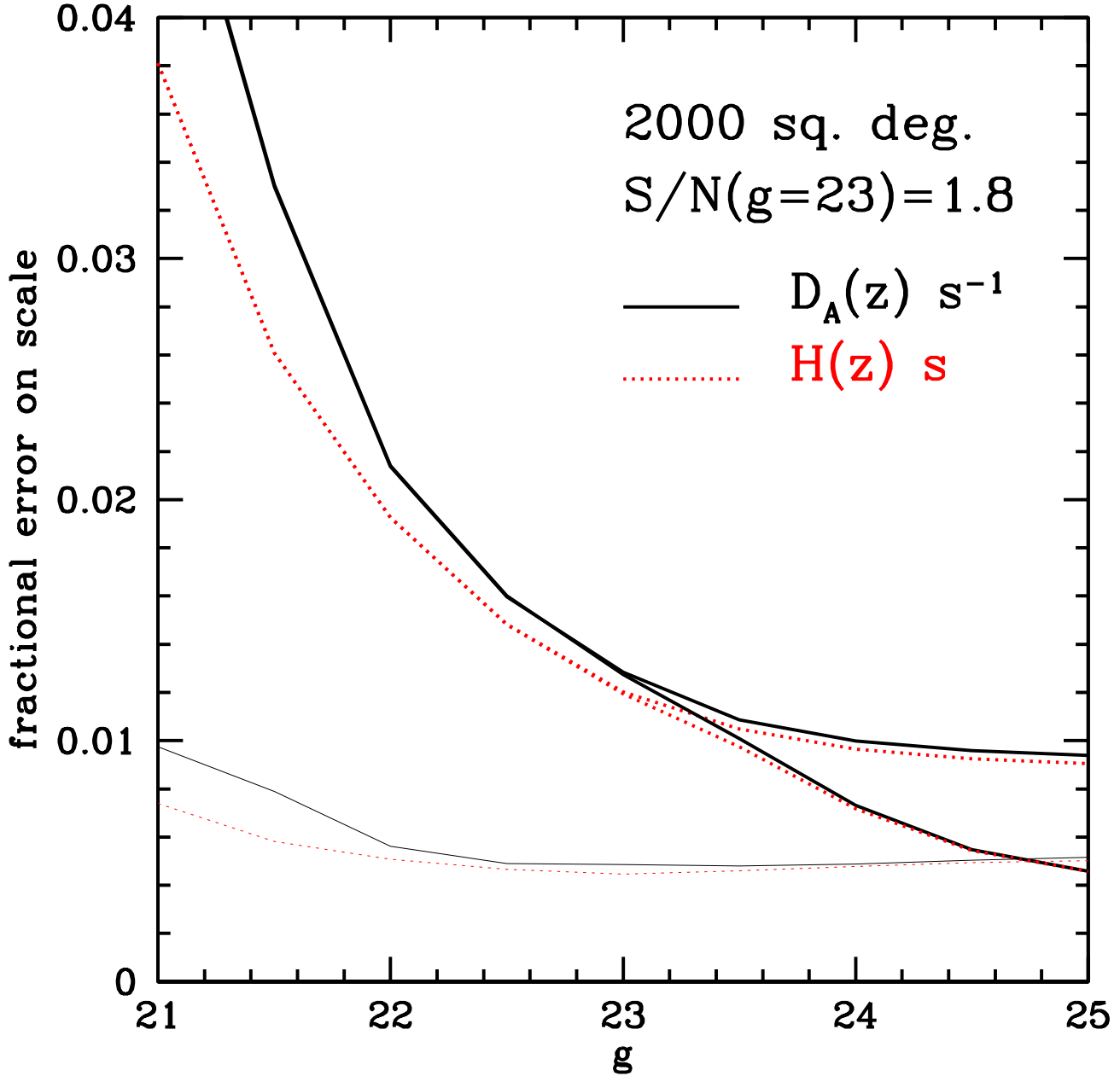}}
\caption{
Upper thick lines show constraints as a function of g magnitude limit on the 
radial and 
transverse BAO scales for a survey similar to the proposed WFMOS low-z galaxy 
survey \citep{2005astro.ph..7457G}, assuming 2000 sq. deg. and R=2000.  
Magnitude limits (21, 22, 23, 24, 25) correspond to (8, 20, 41, 77, 136) 
quasars per sq. deg. for the \cite{2006astro.ph..2569J} luminosity function,
and we assume S/N=(11, 4.5, 1.8, 0.7, 0.3) per \AA.  Lower thick curves add 
spectra from LBGs.   
Thin lines show the completely unrealistic case where we ignore the 
aliasing-like noise power caused by discrete sampling.
}
\label{wfmoswide}
\end{figure}

Figure \ref{wfmoswide} shows the first results, where we assume 
that we are observing the 
\lyaf\ in the wavelength range 3900-5229\AA\ ($2.2<z<3.3$).
The WFMOS galaxy survey \cite{2005astro.ph..7457G} 
is expected to obtain $1\sigma$ errors of 1.0 and 1.2\% on 
$D_A(z)$ and $H(z)$, respectively, at $z\sim 1$, and, 
more relevantly, 1.5 and 
1.8\% at $z\sim 3$ from a separate, similarly costly, high z galaxy survey 
reaching $R<24.5$.  
Our errors depend on the magnitude limit, reaching better than 
1.4\% in both directions for $g<23$. 
The results improve more slowly with increasing magnitude limit
because the faint spectra, at fixed 
observing time, are becoming too noisy to be very useful.  
If very faint quasars and LBGs can be identified, the errors could be
as small as 0.5\%.
Note that the \lyaf\ BAO measurement generally constrains $H(z)$ better than
$D_A(z)$, the opposite trend from galaxies \citep{2005astro.ph..7457G}.

\begin{figure}
\resizebox{\textwidth}{!}{\includegraphics{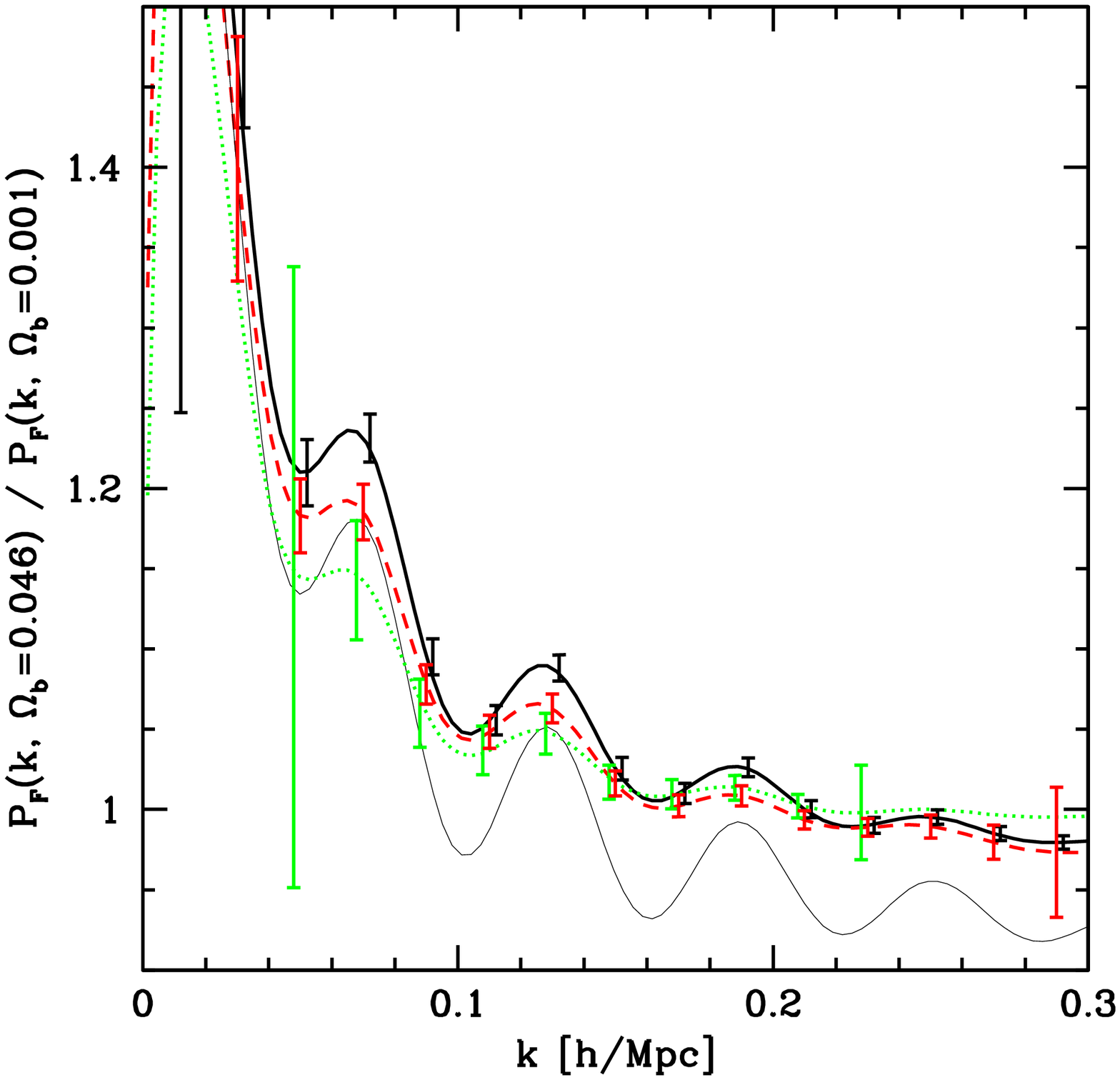}}
\caption{
Error bars show the fractional
error on the power in bands of $\Delta k=0.02 \ihmpc$,  
for the $g<25$ case from Fig.~\ref{wfmoswide} (without LRGs).
Lines show the ratio of \lyaf\ flux power for $\Omega_b=0.0462$ to 
$\Omega_b=0.001$.
Black (solid line, error bars shifted slightly
right) shows $\mu=k_\parallel/k>2/3$, red (dashed) shows 
$1/3<\mu<2/3$, and green (dotted, error bars shifted left) shows $\mu<1/3$.  
For comparison, the thin line shows the ratio of power without aliasing noise 
(renormalized for clarity).
}
\label{bandpower}
\end{figure}

To make the results clearer, we show the obtainable constraints on band power 
measurements
in Fig.~\ref{bandpower} (our standard constraints on the BAO scale do not go 
through this intermediate step).
We see how the inclusion of aliasing-like power from the 
discrete transverse sampling, i.e., the 2nd term in Eq.~(\ref{eqnewpower}), 
dilutes the amplitude of the
BAO features, especially for modes transverse to the line of sight which are
not enhanced by large-scale peculiar velocities.  
The absence of low-k transverse modes can be traced, somewhat 
counter-intuitively, to the fairly large minimum $k_\parallel$ resulting from
the relatively small radial extent of the survey compounded by the loss of 
modes dropped to allow for continuum fitting.
The absence of high k transverse modes is due to the sparse transverse 
sampling.
Spectral noise power is, by construction, always subdominant ($\sim 50$\% of 
the signal power including aliasing power), because quasars faint enough to 
contribute a lot of noise power are discarded by the weighting.

\begin{figure}
\resizebox{\textwidth}{!}{\includegraphics{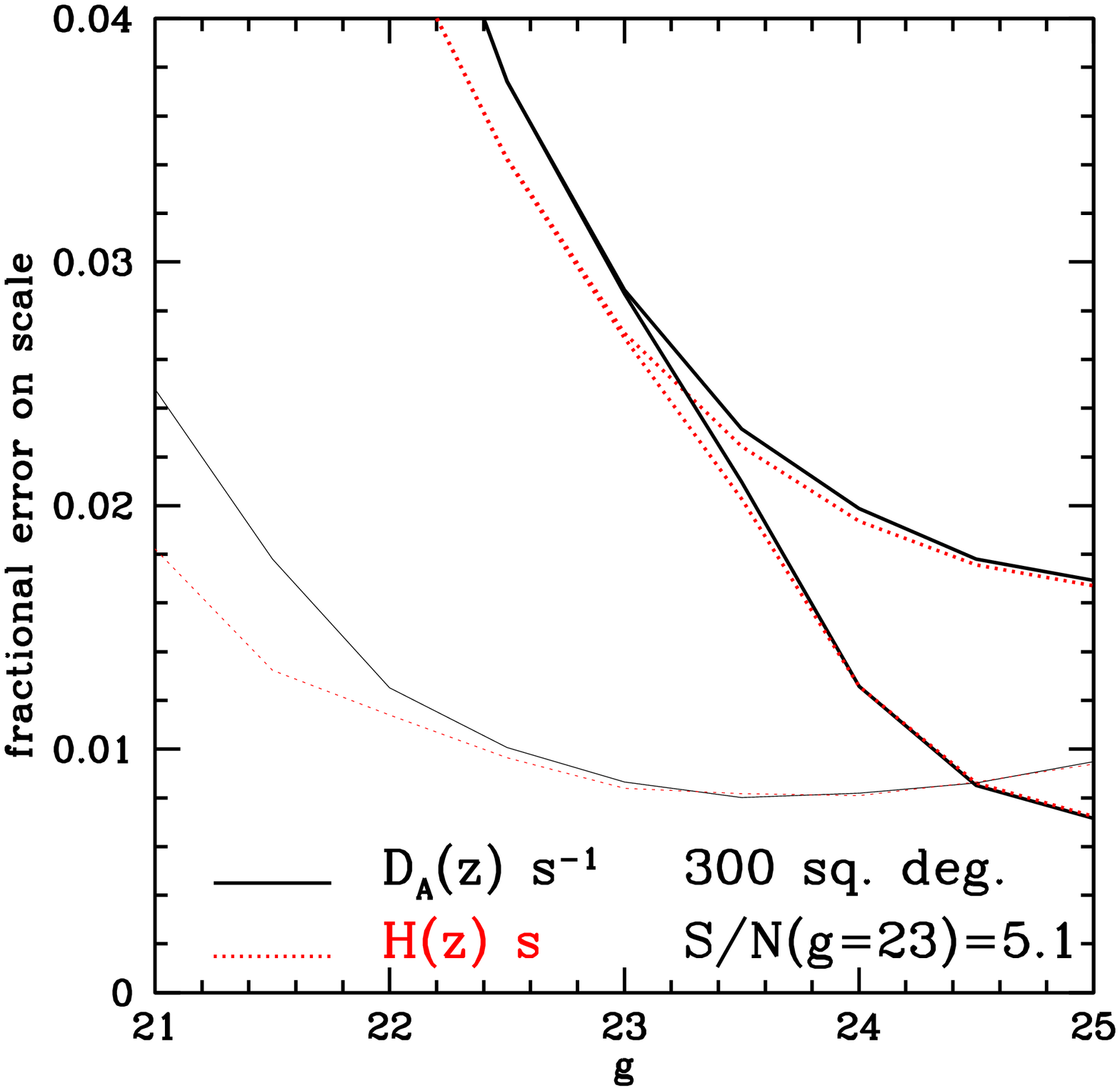}}
\caption{
Similar to Fig.~\ref{wfmoswide}, except covering 300 sq. deg. with longer
exposures, following the high-z galaxy survey proposed in 
\cite{2005astro.ph..7457G}.
g magnitude limits (21, 22, 23, 24, 25) correspond to 
S/N=(32, 13, 5.1, 2.0, 0.8) per \AA. }
\label{wfmosdeep}
\end{figure}

For comparison, in Fig.~\ref{wfmosdeep} we imagine piggy-backing the \lyaf\ 
survey onto the
high-z galaxy survey of \cite{2005astro.ph..7457G}, which is proposed to cover
300 sq. deg. with 240 minute exposures, obtaining errors of 1.8\% on $H(z)$ and
1.5\% on $D_A(z)$ using galaxies (note that the S/N we assume is guided by but
should not be taken as a prediction for WFMOS sensitivity).
We find similar errors to those from galaxies at a similar limiting magnitude, 
but the wider, shallower survey is probably a better option for the \lyaf. 
These figures uncover an important issue for survey planning:  for a 
fixed survey configuration, the
results are quite sensitive to the limiting magnitude, i.e., how bright does
a quasar have to be to be identifiable?   

\begin{figure}
\resizebox{\textwidth}{!}{\includegraphics{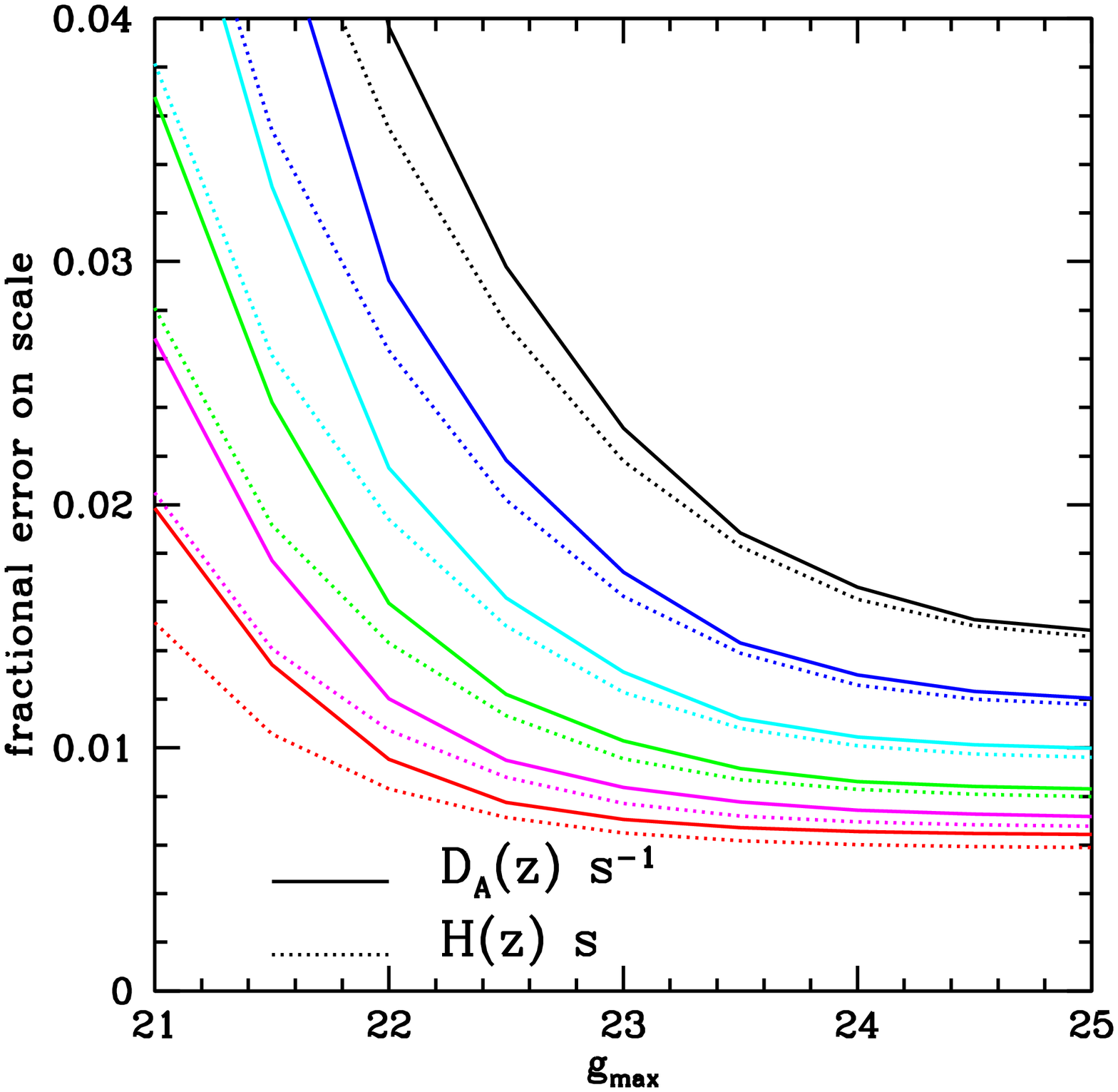}}
\caption{
Similar to Fig.~\ref{wfmoswide}, but investigating the effect of the 
trade-off between noise and area.
From top to bottom (black, blue, cyan, green, magenta, red) we
show, at g=22.5, S/N=(5.7, 4.1, 2.9, 2.0, 1.4, 1.0) per \AA.
The results improve with {\it decreasing} S/N because in each case the area
of the survey is $A=2000~(2.9~N/S)^2$ sq. deg., assuming sky-dominated noise.
Errors scale as $A^{-1/2}$.
}
\label{wfmoswide_scale}
\end{figure}

\begin{figure}
\resizebox{\textwidth}{!}{\includegraphics{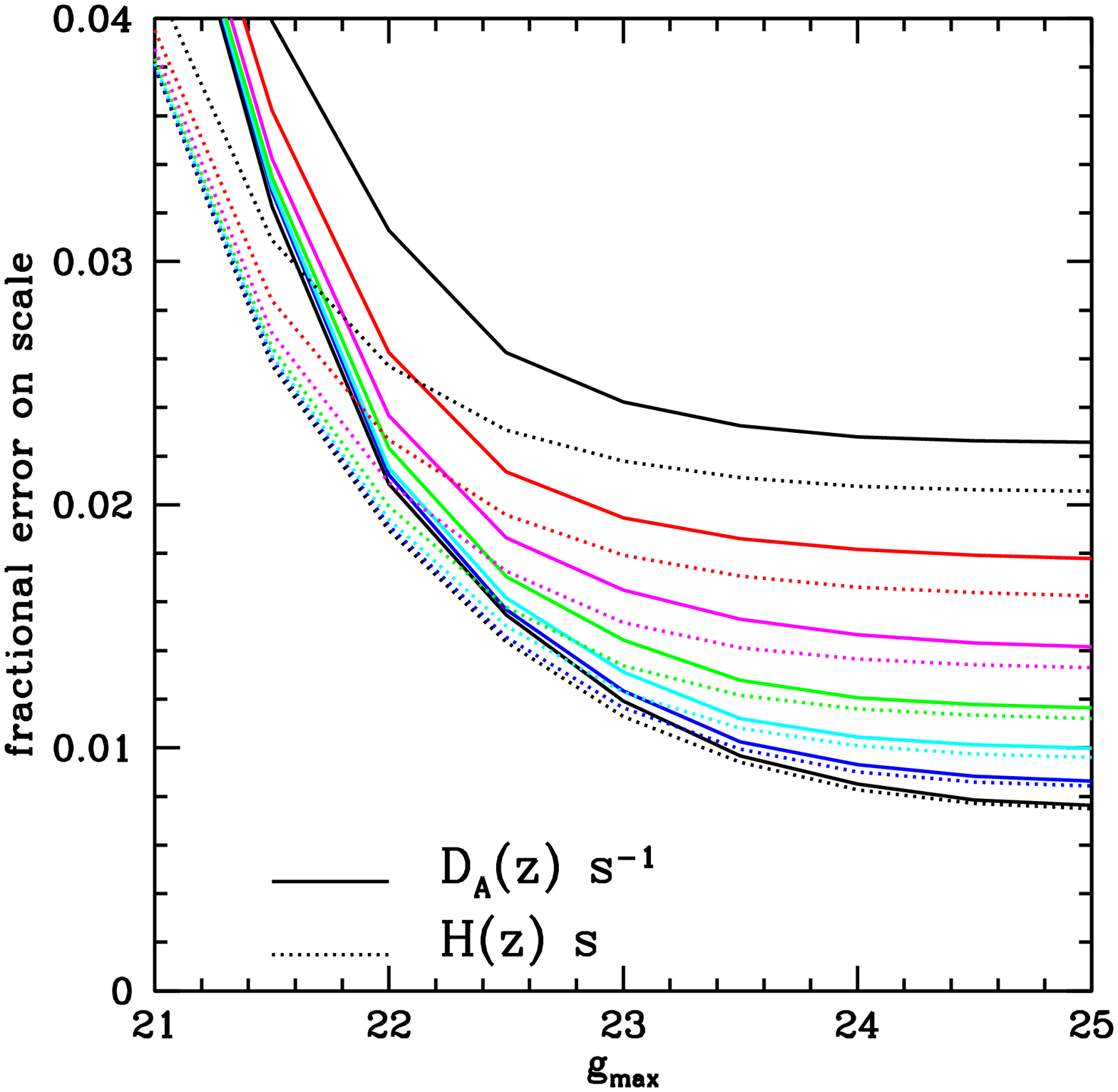}}
\caption{
Similar to Fig.~\ref{wfmoswide}, but investigating the effect of the 
changing the noise level at fixed 2000 sq. deg. survey area. 
From bottom to top (black, blue, cyan, green, magenta, red, black) we
show, at g=22.5, S/N=(5.7, 4.1, 2.9, 2.0, 1.4, 1.0, 0.72) per \AA.
}
\label{wfmoswide_noise}
\end{figure}

We now explore the optimization of the survey.  The errors scale precisely as
$A^{-1/2}$ down to surveys much smaller than the ones we are discussing. 
It seems likely that any 
spectrograph used for this project will have more fibers than necessary, so the
main question is how long to integrate before moving on to gather more area.
Figure \ref{wfmoswide_scale} shows the expected errors as a function of the 
S/N obtained per \AA\ for a g=22.5 quasar, with a corresponding rescaling of
the survey area to keep $A~(S/N)^2$ fixed.  
As in previous figures, we are
including a realistic distribution of magnitudes, i.e., the noise is higher
for fainter quasars and lower for brighter ones.       
We see that a wider, noisier survey is generally superior.  Limitations on 
this progression will be set by telescope overhead and the minimum S/N needed
to identify a quasar.   
We isolate the change with noise level in Fig.~\ref{wfmoswide_noise}.

\begin{figure}
\resizebox{\textwidth}{!}{\includegraphics{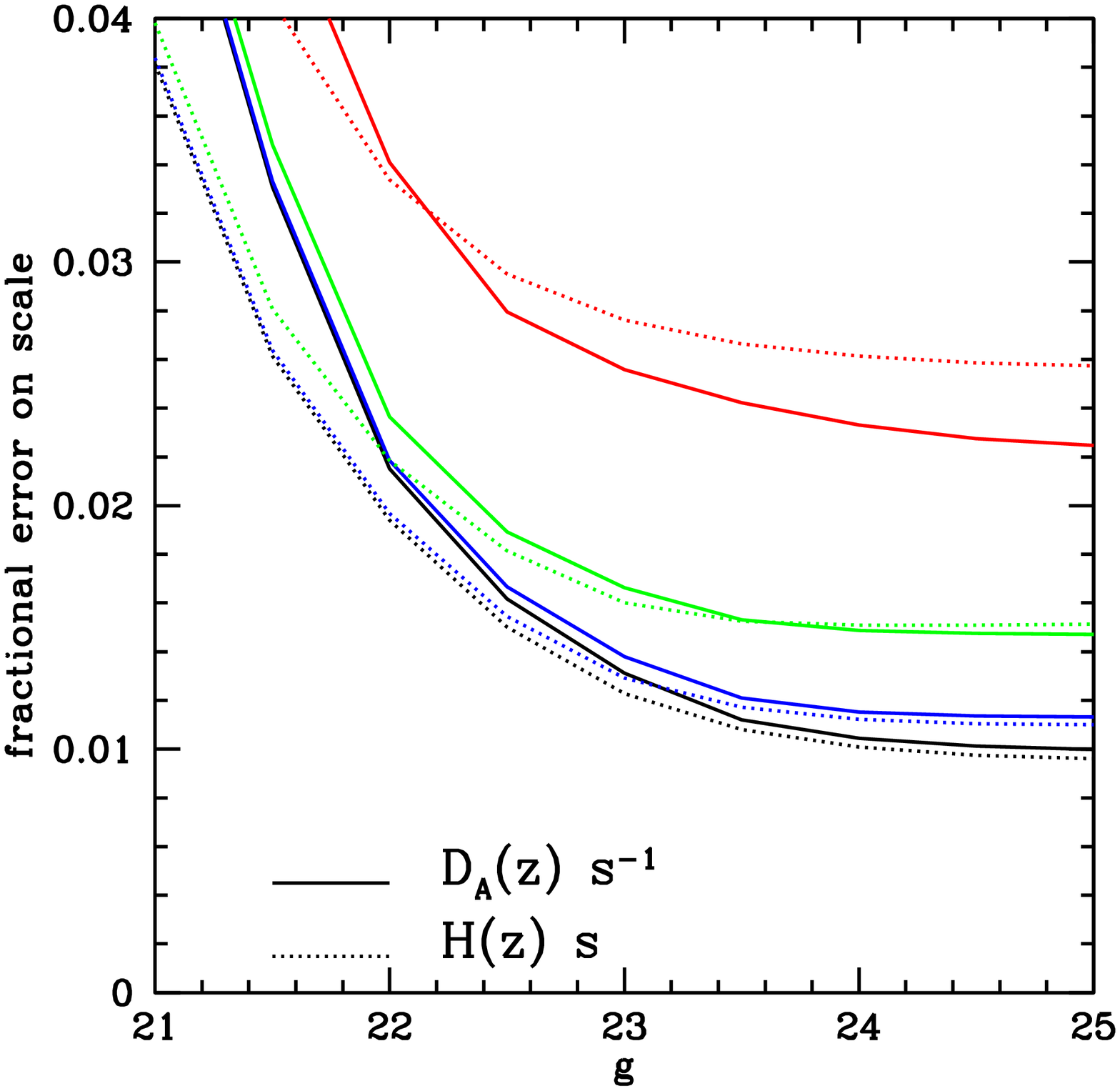}}
\caption{
Similar to Fig.~\ref{wfmoswide}, but investigating the effect of 
changing resolution.
From bottom to top (black, blue, green, red) lines show 
R=(2000, 250, 125, 62.5).
}
\label{wfmoswide_res}
\end{figure}

In Fig.~\ref{wfmoswide_res}, we show the effect of decreasing 
resolution.
Only relatively poor resolution is needed.  R=250 is essentially as good as 
2000, and 125 is not too bad.  These numbers are easy to understand.  At
R=125, the power suppression factor at our FKP weight point, $k=0.07\ihmpc$,
$\mu=0.5$, is only 11\%.  This suppression factor increases very 
quickly with $k_\parallel$; however, the aliasing noise that we show in
Fig.~\ref{bandpower} is itself suppressed by limited resolution, so the 
effect is somewhat less than one might otherwise expect. 

Finally, we perform a few tests to make sure our calculation is robust.  First,
to test our code, we recompute the errors for the 
galaxy survey in \cite{2005astro.ph..7457G}.  
We can do this easily by dropping the aliasing term, setting 
the noise variance in cells defined by the pixel size and mean separation of 
the spectra to the inverse of the mean number of galaxies in a cell, and  
replacing the \lyaf\ flux power spectrum with the galaxy power spectrum. 
We find errors 1.5\% on $D_A(z)$ and 1.6\% on $H(z)$, assuming 
they used $\Omega_m=0.35$, $\sigma_8=0.9$, $b=3.3$, $h=0.65$, $n=1.0$, 
following \cite{2003ApJ...598..720S}.  This compares well with 
\cite{2005astro.ph..7457G}'s 1.5\% and 1.8\%, respectively. 

One approximation we have made is to ignore evolution, i.e., to use the power
spectrum, quasar number density, etc. from the center of the redshift interval
for the full interval.
As a test, we split the sample into two redshift bins, and compute
the Fisher matrix separately for each, including marginalizing over two 
independent sets
of nuisance parameters.  The errors on each bin increase of course, but the 
combination of the two bins actually gives slightly smaller errors than 
the original full Fisher matrix, because of small differences in the averaging.
Not surprisingly, the low redshift half of
the survey gives significantly smaller error bars than the high redshift half 
(the quasar density is higher at fixed apparent magnitude at low $z$).

To test that we are really measuring the BAO feature and not some  
broadband feature in the power spectrum, we run our error computation using
a transfer function with $\Omega_b=0.001$.  We find that the errors for
the survey in Fig.~\ref{wfmoswide} are always greater than 5\%, i.e., our 
measurement is clearly based on the baryonic feature.  Finally, we note that
our results are completely insensitive to removing the marginalization over 
$\beta$ or adding a marginalization over the noise amplitude.

A 30 sq. deg. pilot study reaching $g<21$ with $S/N=11$ at $g=21$ should 
detect baryon oscillations at $\sim 2 \sigma$, in the sense that the Fisher
matrix prediction for a measurement of $\Omega_b$, marginalized over the
nuisance parameters discussed above, predicts an error 0.023 for 
$\Omega_b=0.046$.  Reaching $g<22$ with $S/N>4.5$ would produce a $2.7\sigma$
detection.  A detection of this kind would not be fundamentally interesting;
rather, we give these numbers to indicate the type of survey that would be
needed to produce a good solid measurement of the \lyaf\ power on the 
appropriate scale.  As usual, a survey four times larger with half this 
$S/N$ would produce a better measurement.

\section{Constraints on Dark Energy and Curvature \label{seccosmology}}

In \S\ref{secparametric}, we give projected constraints on the most standard
specific parameterization of the dark energy.
In \S\ref{secnonparametric} we discuss the general usefulness of a BAO 
measurement at $z>2$, in a non-parametric way.  

\subsection{Parametric \label{secparametric} models}

We discuss first the usefulness of a BAO measurement at $z\sim 2.8$ for 
constraining dark energy with equation of state 
$w(a)=w_{0.6}+(0.6-a)~w^\prime$, with $w_{0.6}$ and $w^\prime$ as  
parameters.  We saw above that
the true center of weight of the \lyaf\ survey would probably be a bit lower 
than $z\sim 2.8$ but we find negligible sensitivity of the final 
constraints to the exact \lyaf\ redshift.
The pivot point $a=0.6$ was chosen to make $w_a$ and $w^\prime$ 
roughly
uncorrelated for our scenarios.  We use the figure of merit (FoM) defined by
the Dark Energy Task Force 
\footnote{http://www.hep.net/p5pub/AlbrechtP5March2006.pdf} (DETF) as our 
primary measure of survey value.
The DETF FoM is simply the inverse of the area within the 95\% confidence 
contour in the $w_a-w^\prime$ plane, which has the important 
characteristic of being independent of the chosen pivot point.

We include the projected constraints from the Planck CMB experiment.
We use standard Fisher matrix techniques 
\citep{1999ApJ...518....2E,2002PhRvD..65b3003H}, following 
\cite{Bond:2004rt} in ignoring foregrounds
but using only the 143 GHz channel.  Our results agree well with 
\cite{Bond:2004rt}
when we use similar parameter combinations.  For reproducibility, our results
are given in Table \ref{planckfishmat}.
\begin{table}
\caption{
Assumed values, errors, and error correlations from Planck.
The first row is the assumed value of the parameter, the 
second is the error, and the rest are the correlation matrix.
Very large errors should be interpreted only qualitatively.
\label{planckfishmat}}
\begin{ruledtabular}
\begin{tabular}{lccccccccc}
 & $n_s$ & $w_{0.6}$ & $\Omega_k$ & $\omega_b$ & $\omega_m$ & $\theta_s$ & $\tau$ & $\log_{10}(A)$ & $w^\prime$ \\
\hline
$p$ & 0.963 & -1.00 & 0.00 & 0.0227 & 0.145 & 0.597 & 0.0994 & -8.65 & 0.00 \\
$\sigma$ & 0.0043 & 0.22 & 0.021 & 0.00017 & 0.0015 & 0.00032 & 0.0046 & 0.0038 & 4.1 \\
$n_s$ & 1.000 & -0.078 & 0.034 & 0.594 & -0.842 & 0.294 & 0.320 & -0.075 & -0.119 \\
$w_{0.6}$ & -0.078 & 1.000 & 0.577 & 0.058 & -0.004 & -0.104 & 0.034 & 0.045 & -0.006 \\
$\Omega_k$ & 0.034 & 0.577 & 1.000 & 0.161 & -0.117 & 0.192 & 0.180 & 0.146 & 0.315 \\
$\omega_b$ & 0.594 & 0.058 & 0.161 & 1.000 & -0.645 & 0.352 & 0.252 & 0.000 & -0.019 \\
$\omega_m$ & -0.842 & -0.004 & -0.117 & -0.645 & 1.000 & -0.282 & -0.308 & 0.110 & 0.069 \\
$\theta_s$ & 0.294 & -0.104 & 0.192 & 0.352 & -0.282 & 1.000 & 0.079 & -0.010 & 0.684 \\
$\tau$ & 0.320 & 0.034 & 0.180 & 0.252 & -0.308 & 0.079 & 1.000 & 0.899 & -0.059 \\
$\log_{10}(A)$ & -0.075 & 0.045 & 0.146 & 0.000 & 0.110 & -0.010 & 0.899 & 1.000 & -0.018 \\
$w^\prime$ & -0.119 & -0.006 & 0.315 & -0.019 & 0.069 & 0.684 & -0.059 & -0.018 & 1.000 \\
\end{tabular}
\end{ruledtabular}
\end{table}
We see that the CMB alone only weakly constrains dark energy (note here that
the Fisher matrix technique may not give reliable errors in cases where the
errors are large, but this should not be a problem once other data constrains
these degenerate directions).

Table \ref{contable} shows the FoM and constraints on $w_{0.6}$, 
$w^\prime$, and 
sometimes $\Omega_k$, for several different combinations of data.
\begin{table}
\caption{
Errors on $w_{0.6}$, $w'$, and $\Omega_k$ 
[with $w(a)=w_{0.6}+(0.6-a)~w^\prime$] for different data combinations, along 
with
the DETF figure of merit (FoM), defined to be the inverse area inside the 
$2-\sigma$ contours in the $w_{0.6}-w'$ plane.
BAO errors are 5.8\% radial and 5.2\% transverse at $z=0.3$, and 1\% in both
directions at $z=2.8$.  The $z\sim 1$ constraint is a set of points in the
range $0.6<z<1.2$, with combined precision equal to the $z=2.8$ constraint (see
text).
\label{contable}}
\begin{ruledtabular}
\begin{tabular}{ccccccccc}
 & BAO & BAO & BAO & & & & &\\
Planck & $z=0.3$ & $z\sim1$ & 
$z=2.8$ & FoM & $\sigma_{w_{0.6}}$ & $\sigma_{w'}$ 
& $\sigma_{\Omega_k}$ \\
\hline
Y & Y & N & N & 0.16 & 0.211 & 2.37 & 0.0100 \\
Y & Y & Y & N & 0.75 & 0.091 & 1.48 & 0.0038 \\
Y & Y & N & Y & 0.52 & 0.127 & 1.36 & 0.0031 \\
Y & Y & Y & Y & 1.40 & 0.068 & 0.84 & 0.0022 \\
Y & Y & N & N & 0.27 & 0.130 & 2.30 & --- \\
Y & Y & Y & N & 1.31 & 0.074 & 0.87 & --- \\
Y & Y & N & Y & 0.54 & 0.122 & 1.35 & --- \\
Y & Y & Y & Y & 1.61 & 0.067 & 0.74 & --- \\
\end{tabular}
\end{ruledtabular}
\end{table}
To avoid sensitivity to broad-band power, we use separate (independent) slope 
and amplitude parameters for the CMB and \lyaf.
We use three different BAO constraints:  First, we always assume that SDSS will 
produce a 5.8\% error on 
the radial scale at $z=0.3$, $v_s(z=0.3)$, and 5.2\% error on the transverse 
scale, $\theta_s(z=0.3)$ \citep{2003ApJ...598..720S}.  We optionally
add a 1\% constraint on both distance scales at $z=2.8$.  Finally, we 
optionally include a $\sim 1$\% constraint at $z\sim 1$.  To avoid artificially
degrading the value of a $z\sim 1$ survey, we spread this measurement over 
points at $z=0.6, 0.8, 1.0$, and 1.2, using the relative errors from
\cite{2003ApJ...598..720S} but modifying the overall normalization to make the
combined error from the eight points $0.7$\%, i.e., the same total precision 
as the two 1\% $z=2.8$ measurements.  We choose 
a simple 1\% because different survey configurations can lead to many different
combinations of errors.  The point here is primarily to study the general 
usefulness of constraints at different redshifts, not any specific survey.

We see, as expected, that in the presence of Planck constraints the
measurement at $z=2.8$ is less valuable for constraining dark energy than the 
measurement at 
$z\sim1$; however, if $\Omega_k$ is allowed to vary, adding the $z=2.8$ 
measurement nearly doubles the DETF FoM.  This improvement is larger than 
the affect of doubling the size of the $z\sim 1$ survey.  
The higher $z$ measurement is actually slightly more valuable than $z\sim 1$ 
for constraining $\Omega_k$.
If we assume flatness the improvement is more modest, although not 
completely negligible.
If the CMB is taken out of play for some reason, the $z=2.8$ measurement 
becomes more valuable than $z=1$, although in that case both are really needed 
to provide an interesting measurement.

We note that the commonly used approximation that the Planck constraint can
be represented by the constraints on $\theta_s$ (0.04\%, or perfectly 
known) and $\omega_m$ (1\%) alone, treated as independent of each other and 
other parameters, works very well, in the sense that the resulting FoM agrees 
with the full-Fisher matrix version to better than 10\%.  

\subsection{Non-parametric \label{secnonparametric}}

We will next consider the measurement of the acoustic scale at $z\gtrsim2$
in the context of testing the flat $\Lambda$CDM model.  We will expand
in small perturbations around the fiducial model, working to lowest order
in the non-constant dark energy and curvature.  
Hence, we write $H^2 = (8 \pi G/3)(\rho_\fid + \rho_X)+\Omega_K H_0^2(1+z)^2$, 
where $\rho_\fid$ is the sum of the matter, radiation, and
cosmological constant contributions and $\rho_X$ includes the dark
energy density that differs from the $z=0$ value.  
We want to test whether $\rho_X=0$
or not, keeping in mind the uncertainties in $\rho_\fid$ and curvature.

\subsubsection{Transverse Scale}

In the transverse direction, we measure $\ell_z = \theta_s(z)^{-1}=
(1+z) D_A(z)/s$.  For example,
we measure $\ell_{1089}$ to wonderful accuracy (0.35\% with 3-year WMAP,
$\ll0.1\%$ with Planck).  Now we consider a measurement at $z\approx2$
and construct $\ell_{1089}-\ell_z$ as a way to isolate the new information
beyond that available in the CMB.
To lowest order in curvature, the angular diameter distance 
\begin{equation}
(1+z) D_A = r + \frac{\Omega_K}{6} \frac{H_0^2 r^3}{c^2}.
\end{equation}
This then yields
\begin{eqnarray}
\ell_{1089}-\ell_z 
&=& \frac{1090 D_A(1089) - (1+z) D_A(z)}{s}
\approx \frac{r(1089)-r(z)}{s} + \frac{\Omega_K H_0^2}{6s c^2}\left[r^3(1089)-r^3(z)\right]
\nonumber \\
&=& \frac{1}{s} \int_z^{1089} \frac{c\;dz}{H}
+ \frac{\Omega_K H_0^2}{6s c^2} \left[ r^2(1089)+r(1089) r(z) + r^2(z) \right]
    \int_z^{1089} \frac{c\;dz}{H_\fid} \nonumber \\
&\approx& \frac{1}{s} 
	\left[ 1 + \frac{\Omega_K}{6}\frac{H_0^2 r^2(1089)}{c^2} \left( 1 + d + d^2\right)\right]
	\int_z^{1089} \frac{c\;dz}{H_\fid} 
	- \frac{1}{2s} \int_z^{1089} \frac{c\;dz }{H_\fid} 
	\left(\frac{\rho_X}{\rho_\fid} + \frac{3\Omega_K H_0^2(1+z)^2}{8\pi G \rho_\fid}\right)
\label{elldef}
\end{eqnarray}
where we've define $d = r(z)/r(1089)$ and 
assumed that $|\Omega_K|\ll1$ and $\rho_X\ll\rho_\fid$
(suitable for $z\gtrsim2$).
The latter means that $\rho_X/\rho_\fid = \Omega_X$.
Keeping only lowest order in the perturbations around the fiducial model, we also have
\begin{equation}
\frac{3\Omega_K H_0^2(1+z)^2}{8\pi G \rho_\fid} = 
\frac{\Omega_K(1+z)^2}{\Omega_m (1+z)^3 + \Omega_\Lambda} 
\approx \frac{\Omega_K}{\Omega_m (1+z)} \left[1-2\lambda+O(\lambda^2)\right]
\end{equation}
where $\lambda = \Omega_\Lambda/2\Omega_m(1+z)^3$.  We can drop the 
higher orders in $\lambda$ when working at higher redshift.


We now define a weighted average of $\Omega_X$
\begin{equation}
\bar\Omega_X \equiv \frac{ \int_z^{1089} \frac{c\;dz}{ H_\fid}\Omega_X 
	}{ \int_z^{1089} \frac{c\;dz}{ H_\fid} }
\end{equation}
and rearrange to find 
\begin{equation}
\bar\Omega_X = 2 + \frac{2 s}{ \int_z^{1089} \frac{c\;dz}{ H_\fid}} 
\left(\ell_z - \ell_{1089}\right)
+ \Omega_K \left[ \frac{H_0}{3 c^2} r^2(1089)(1+d+d^2) - 
			\frac{1-\frac{6\lambda}{7}}{3 \Omega_m (1+z)} \right].
\end{equation}
The last term comes from doing the integral for the curvature term
in (\ref{elldef}).  For standard cosmologies, the coefficient in
square brackets is about 5 for $z=2$-3.  The middle term must be
$-2$ in the fiducial model, because the result cancels to $\bar\Omega_X=0$.

The question is now how accurately we can measure $\bar\Omega_X$
in light of the uncertainties on the various terms on the
right-hand side of the equation.  In particular, we want to know
how the uncertainties in $\Omega_m h^2$, $\Omega_m$, and $\Omega_\Lambda$
enter (i.e., how well specified the baseline model is).
It is useful to rearrange the middle term as
\begin{equation}
2 \left[s(\Omega_mH_0^2)^{1/4}\right] (\Omega_m H_0^2)^{1/4}
\left[\sqrt{\Omega_m H_0^2} \int_z^{1089} \frac{c\;dz}{ H_\fid}\right]^{-1} 
\left(\ell_z - \ell_{1089}\right)
\end{equation}
The combination $s(\Omega_m H_0^2)^{1/4}$ is picked to cancel out most
of the dependence of $s$ on $\Omega_m H_0^2$.
We can manipulate the integral as
\begin{equation}
\int_z^{1089} \frac{c\;dz}{ H_\fid}
= \frac{c}{ \sqrt{\Omega_m H_0^2}} \int_z^{1089} 
\frac{dz }{ \sqrt{(1+z)^3 + (\Omega_r/\Omega_m) (1+z)^4}}
\left[1+\frac{\Omega_\Lambda}{ \Omega_m} (1+z)^{-3}\right]^{-1/2}~.
\end{equation}
If we expand the square bracket terms and neglect the radiation
term when integrating the terms with $\Lambda$, then we find
\begin{equation}
\int_z^{1089} \frac{c\;dz}{ H_\fid}
= \frac{c}{ \sqrt{\Omega_m H_0^2}} \left[
\int_z^{1089} \frac{dz }{ \sqrt{(1+z)^3 + (\Omega_r/\Omega_m) (1+z)^4}}
- \frac{2}{\sqrt{1+z}}\left(\frac{\lambda}{ 7} - \frac{3\lambda^2}{ 26} + 
\ldots\right)\right]
\end{equation}
\begin{equation}
= \frac{2c}{ \sqrt{\Omega_m H_0^2}} \left[
\left.\sqrt{a+a_{eq}}\right|^z_{1089}
- \frac{1}{\sqrt{1+z}} \left(\frac{\lambda}{ 7} - \frac{3\lambda^2}{ 26} + 
\ldots\right)\right]
\end{equation}
where $a=1/(1+z)$. 
With this, we find
\begin{eqnarray}
\bar\Omega_X &=& 2 + \frac{\sqrt{1+z}}{ c}
\left[s(\Omega_mH_0^2)^{1/4}\right] (\Omega_m H_0^2)^{1/4}
\left(\ell_z - \ell_{1089}\right)
\left[1 - \sqrt{(1+z)\left(\frac{1}{1090}+\frac{1}{ z_{eq}}\right)}
- \frac{\lambda}{ 7} + \frac{3\lambda^2}{ 26} + O(\lambda^3)
\right]^{-1} \nonumber \\
&& + \Omega_K \left[ \frac{H_0}{3 c^2} r^2(1089)(1+d+d^2) - 
			\frac{1-\frac{6\lambda}{7} + O(\lambda^2)}{\Omega_m (1+z)} \right].
\end{eqnarray}

Now we can look at the errors in these terms.  We denote $\sigma(x)$
as the standard deviation of $x$.
The error in $\Omega_m h^2$ is expected to be below 2\% with Planck data.
That means that the fractional error in $(\Omega_m H_0^2)^{1/4}$ is 
below 0.5\%.
The quantity $s(\Omega_m H^2)^{1/4}$ essentially depends only on
$\Omega_bh^2$, but this is to the $1/8$ power, so the error
will be below 0.1\%.
For $\lambda$, 
\begin{equation}
\sigma(\lambda) = \frac{1}{2(1+z)^3\Omega_m} \frac{\sigma(\Omega_m)}{\Omega_m}.
\end{equation}
As we care about $\lambda/7$, the value ends up being about
$\sigma(\lambda/7) = \sigma(\Omega_m) (1+z)^{-3}$.  With today's cosmological
constraints (e.g., \cite{2005ApJ...633..560E,2006astro.ph..4335S}), 
this error is about 0.1\% at $z=2$; of course, this will shrink in 
the future.  In other words, our low redshift
data constrains $\Lambda$ well enough that the uncertainties in the 
extrapolation to $z>2$ are tiny.  The contribution of $\lambda$ in 
the curvature term is smaller yet; we will drop these.
The errors in the radiation term $z_{eq}$ are very small.

This leaves the fractional error in $\ell_z-\ell_{1089}$.  The
error in $\ell_{1089}$ will be very small with Planck, so we neglect
it.  Then
\begin{equation}
\frac{\sigma(\ell_z-\ell_{1089})}{ \ell_z-\ell_{1089}}
= \frac{\sigma(\ell_z)}{ \ell_z} \frac{1}{ \ell_{1089}/\ell_z -1 }
\end{equation}
The quantity $\ell_{1089}/\ell_z-1$ is 1.7 for $z=2$ and 1.2 for $z=3$.

Hence, the error in $\bar\Omega_X$ is dominated by the quadrature sum
of half the fractional error in $\Omega_mh^2$ ($\lesssim1\%$ for Planck),
the fractional error in $\ell_z$ times $2/(\ell_{1089}/\ell_z-1)$
(which is 1--1.7, depending on redshift), and five times the uncertainty
in the curvature $\Omega_K$.  Of course, one might opt to move the
curvature to the other side of the ledger and constrain $\bar\Omega_X - 5\Omega_K$.

It is a good approximation to think of $\bar\Omega_X$ as
\begin{equation}
\bar\Omega_X \approx \frac{
\int_{a_{eq}}^a d\ln a\;a^{1/2} \Omega_X
}{ \int_{a_{eq}}^a d\ln a\;a^{1/2} }
\end{equation}
One might compare this to the effect of anomalous dark energy on
the growth of structure, which enters as a suppression of growth
approximately as a factor
\begin{equation}
1 - \frac{3}{ 5}\int_{a_{eq}}^a d\ln a \;\Omega_X
\end{equation}
The weightings differ by $a^{1/2}$.


\subsubsection{Radial Scale}

Next, we turn to the radial acoustic scale.  Here we are measuring
$v_s(z) = s H(z)/(1+z)$ at a given $z$.  Performing the same expansion, we have
\begin{equation}
v_s(z) = s \frac{H_\fid(z)}{1+z} \left( 1 + \frac{\rho_X}{ 2\rho_\fid} + 
\frac{3\Omega_K H_0^2(1+z)^2}{8\pi G \rho_\fid}\right)
= s~(1+z)^{1/2}\sqrt{\Omega_m H_0^2} \left[ 1 + \frac{\Omega_\Lambda }{
\Omega_m(1+z)^3}\right]^{1/2} \left(1+\frac{\Omega_X}{2} 
+ \frac{1}{2}\frac{3\Omega_K H_0^2(1+z)^2}{8\pi G \rho_\fid}
\right)
\end{equation}
If we expand in powers of $\lambda$, we get
\begin{equation}
v_s(z) = \left[s(\Omega_mH_0^2)^{1/4}\right] (1+z)^{1/2}(\Omega_m H_0^2)^{1/4}
\left[1+\lambda-\frac{\lambda^2}{ 2} + O(\lambda^3)\right] 
\left[1+\frac{\Omega_X}{2} + \frac{\Omega_K}{2 \Omega_m(1+z)}(1-2\lambda+O(\lambda^2))\right]
\end{equation}
Solving for $\Omega_X$, we have
\begin{equation}
\Omega_X = -2 + 2~v_s(z)~(1+z)^{-1/2} \left[s(\Omega_mH_0^2)^{1/4}\right]^{-1} 
(\Omega_m H_0^2)^{-1/4}
\left[1+\lambda-\frac{\lambda^2}{ 2} + O(\lambda^3)\right]^{-1}
- \frac{\Omega_K}{\Omega_m(1+z)}(1-2\lambda+O(\lambda^2))
\end{equation}
Again, the middle term must be $2$ in the fiducial model.

Looking at the error budget, again we expect to know $(\Omega_m H_0^2)^{1/4}$
to better than 0.5\%.  The error in $\lambda$ is about $\sigma(\Omega_m)/4$
at $z=2$, which is already about 0.7\% today and should drop toward 0.1\%.
The coefficient of $\Omega_K$ will be about unity for a survey at $z\approx 3$.
So the error in $\Omega_X$ is about the quadrature sum of half the
fractional error in $\Omega_m H_0^2$ (somewhat below 1\%), twice 
the fractional error in the acoustic scale $v_s(z)$, and the error
on $\Omega_K$.

\subsubsection{Measurement Goals}

It is not clear what quantitative goal one wants to set for
the measurement of $\Omega_X$ or $\bar\Omega_X$.  One goal is to simply
detect the cosmological constant at high redshift, i.e., to exclude 
$\Omega_X(z) = -\Omega_\Lambda(z)$.  When one averages this model over 
redshift, one gets $\bar\Omega_X = \Omega_\Lambda/7\Omega_m (1+z)^3$.
This is only about 1\% at $z=2$.  Hence, it is very hard to detect
the absence of high-redshift dark energy using the transverse distance
scale.  However, it is fairly easy with the radial scale: 
$\Omega_X = -\Omega_\Lambda/\Omega_m(1+z)^3$ is about 10\% at $z=2$
and 4\% at $z=3$.  A 1\% measurement of $H(z)$ at $z=2$ should
detect the dark energy at 3 to 5 $\sigma$, depending on whether the 
error bars on $\Omega_m h^2$ are 2\% or somewhat better.  

Of course, extra dark energy can be detected too.  The 1\% measurement
of $H(z)$ would bound (at 3-5 $\sigma$) the energy density of the dark
energy at $z=2$ to be within a factor of two of its low redshift value.

The above statements
assumes a flat cosmology, but the sensitivity of the transverse scale
to curvature is 5 times larger than that of the radial scale.  Hence,
one can use the transverse acoustic scale to control the curvature to
levels well smaller than the errors on the $\Omega_X$ measurement from the 
radial acoustic scale; this assumes that $\bar\Omega_X\lesssim \Omega_X$.
Putting this another way, one can mix the transverse and radial scale 
to produce a curvature-independent measurement of a combination 
$\Omega_X + \bar\Omega_X/5$, the coefficient being appropriate 
to the redshifts considered here,
Many dark energy models will not
yield zero for this combination; this would require that $\Omega_X$ is 
larger at redshifts above the survey redshift and of the opposite sign.
Hence, one can perform a reasonably generic search for deviations from
the cosmological constant at the survey redshift.  

Whether this is interesting depends considerably on one's model for
dark energy.  The $w=w_0+w_a (1-a)$ model has the unfortunate property
of demanding that there are no changes in dark energy at high redshift
that aren't heralded with even larger changes at low redshifts.  If
one relaxes that assumption it is certainly the case that we don't
know the dark energy density at $z=2$ to 30\% rms.

Another question is whether there is a steady value of $\Omega_X$ at
high redshift, as tracker models would have \cite{1999PhRvL..82..896Z}.  
Here, $\bar\Omega_X$ 
is reasonably close to $\Omega_X$, and one could reach a error bar
of about 1\%, assuming a flat cosmology.  
One may compare this to constraint available from measuring
the growth of structure relative to the CMB. 
Here, one is limited by the optical depth to $z=1000$.  With Planck, 
this may be measured as well as 0.5\%, so if one could measure the growth function to that 
level, one could measure a value of $\Omega_X$ in the 0.1-0.2\%
range (1-$\sigma$).  This method, however, is at least partially 
degenerate with the suppression caused by non-zero neutrino mass
\cite{1998PhRvL..80.5255H}.
A suppression of 0.5\% corresponds to about 0.015 eV, already
well into the region indicated by the atmospheric neutrino results.
Hence, the degeneracy between neutrino mass will be an issue
as one tries to push constraints on $\Omega_X$ below 1\%.

Looking beyond the scope of the surveys discussed in this paper,
the full-sky cosmic variance limit on the acoustic scale at these
redshifts is superb.  At this point, the uncertainty on $\Omega_m h^2$
from CMB experiments such as Planck will produce sufficient 
uncertainty in the acoustic scale as to dominate the error budget
on $\Omega_X$ or $\bar\Omega_X$.  From a statistical point of view,
such BAO surveys themselves could measure $\Omega_m h^2$ better than 
the CMB, thereby restoring the precision in the sound horizon, but
the systematics in broad-band power measurement from low-redshift
surveys will surely be worse than those in the CMB.  Another option
is to form a combination $\Omega_X+\bar\Omega_X$ that is 
independent of $\Omega_m h^2$.  This combination, however, is not 
independent of curvature; essentially one is using the angular acoustic
scale in the CMB to calibrate $\Omega_m h^2$.  Nevertheless, this is
a simple example of the idea that the BAO measurement of $D_A(z)$
and $H(z)$ does produce an internal cross-check, namely that $D_A$
is an integral of $H(z)$, that can be used to eliminate certain nuisance
parameters.

\section{Conclusions}

As a probe of dark energy, baryonic acoustic oscillation features have a 
significant advantage over standard candles such as supernovas in that 
we have a very clear theory describing them.  
Even the imperfection associated with our lack of understanding of exactly
how galaxies populate dark matter halos does not present a significant
problem, because the scale of halos is well separated from the scale of the BAO.
The \lyaf\ extends this advantage in that we can perform something 
relatively close to a computation from first-principles.  

In this paper we have computed the statistical power that can be expected from 
a large, three-dimensional \lyaf\ survey probing the BAO 
scale.  
If the Universe is not assumed to be flat, a measurement of the BAO
scale at $z\sim 3$ provides a interestingly large improvement in the constraints
on dark energy, as quantified in \S\ref{seccosmology} using the Dark Energy 
Task Force figure of merit.
A range of combinations of area and magnitude limit could lead to a 
viable survey, as described in \S\ref{secresults}.
Resolution and signal-to-noise ratio requirements are modest.  It will probably
be most efficient to piggy-back \lyaf\ surveys on top of lower redshift galaxy 
surveys.  

As with other probes, the \lyaf\ version of the BAO measurement should be 
robust against systematic errors.
Long-range effects in the UV ionizing background
will not produce preferred scales and can't mimic the acoustic peak.
The poorly modeled highest density absorption (e.g., DLAs) will not produce
a BAO signal either. 
Continuum features could possibly create a bump on the appropriate scale, but
these should be easy to control because they would be associated with fixed 
quasar rest wavelength ranges, unlike the BAO feature.  In the worst case we
would lose only a small fraction of pixel-pairs by completely ignoring 
correlation between pixels in the same spectrum.
Since we would be using multiobject spectrograph techniques, fluxing errors 
would create
density variations as a function of redshift for all objects in a
region.  This creates fake large-scale power; however,
one can project this purely radial power out with minimal information
loss \citep{1998ApJ...503..492S,1998ApJ...499..555T}.

The ultimate \lyaf\ survey could include faint Lyman break galaxies to obtain a
higher density of probes.  However, getting a redshift for these galaxies is 
not trivial.
For the 75\% that don't have Lya emission, one needs to integrate down
to the continuum, which will require higher S/N than to obtain the redshift
of a quasar QSO.  At this point we also know less about possible relevant 
variations in galaxy continua.

In the short term, a pilot study aimed at simply detecting the BAO feature in 
the \lyaf\ would be very desirable.  The bare minimum requirements are roughly 
$>30$ square degrees with a magnitude limit $g>21$.

\acknowledgements

DJE is supported by NSF AST-0407200 and by an Alfred P.\ Sloan Research 
Fellowship.
Some computations were performed on CITA's Mckenzie cluster which was funded by
the Canada Foundation for Innovation and the Ontario Innovation Trust 
\citep{2003astro.ph..5109D}.

\bibliography{cosmo,cosmo_preprints}

\end{document}